\documentclass[twocolumn]{article}
\usepackage{graphicx}
\usepackage{url}
\usepackage{amsmath}
\usepackage{amssymb}
\usepackage{amsfonts}
\usepackage{color}
\usepackage[utf8]{inputenc}
\usepackage[T1]{fontenc}
\usepackage{float}
\usepackage{subcaption}
\usepackage{caption}
\usepackage{booktabs}
\usepackage{geometry}
\usepackage{multirow}
\usepackage{hyperref}
\usepackage{authblk}
\usepackage{tikz}
\usetikzlibrary{shapes.geometric,arrows.meta,positioning,fit,backgrounds,calc}

\title{Suppressing Domain-Specific Hallucination in Construction LLMs:\\
A Knowledge Graph Foundation for GraphRAG and QLoRA\\
on River and Sediment Control Technical Standards}

\author[1]{Takato Yasuno}

\date{}

\begin{document}
\maketitle

\begin{abstract}

This paper addresses the challenge of answering technical questions derived from Japan's
\textbf{River and Sediment Control Technical Standards}---a multi-volume regulatory document
covering survey, planning, design, and maintenance of river levees, dams, and sabo (torrent
control) structures---using open-source large language models running entirely on local hardware.

We implement and evaluate three complementary approaches:
\textbf{Case~A} (plain 20B LLM baseline),
\textbf{Case~B} (8B LLM with QLoRA domain fine-tuning on 715 graph-derived QA pairs), and
\textbf{Case~C} (20B LLM augmented with a Neo4j knowledge graph via GraphRAG).
All three cases use the Swallow series of Japanese-adapted LLMs and are evaluated on a
100-question benchmark spanning 8 technical categories, judged automatically by an
independent LLM (Qwen2.5-14B, score 0--3).

The key finding is a \textbf{performance inversion}: the 8B QLoRA fine-tuned model (Case~B)
achieves a judge average of \textbf{2.92/3} --- surpassing both the 20B plain baseline
(Case~A: 2.29/3, $+$0.63) and the 20B GraphRAG approach (Case~C: 2.62/3, $+$0.30) ---
while running at \textbf{3$\times$ faster} latency (14.2~s vs.\ 42.2~s for Case~A).
GraphRAG provides moderate gains ($+$0.33 over baseline) but is outperformed by
domain-specific fine-tuning in both quality and efficiency.

We document the full engineering pipeline, including knowledge graph construction
(200 nodes, 268 relations), QLoRA training data generation from Neo4j relations,
training on a single RTX~4060~Ti (16~GB VRAM) using \texttt{unsloth}, GGUF Q4\_K\_M quantisation
and Ollama deployment, and the graph retrieval and re-ranking design.
High-level engineering lessons are distilled in the main body;
implementation pitfalls and toolchain details are documented in Supplementary Materials.

\end{abstract}

\noindent
\textbf{Keywords:} GraphRAG, QLoRA, Knowledge Graph, LLM Fine-Tuning,
River Engineering, Technical Standards, LLM-as-Judge, Japanese NLP,
Open-Source LLM, Domain Adaptation, Domain-specific Hallucination.

\section{Introduction}
\label{sec:intro}

Japan's \emph{River and Sediment Control Technical Standards} (\textit{Kasen Sabo Gijutsu Kijun}),
published by the Ministry of Land, Infrastructure, Transport and Tourism (MLIT),
comprise multi-volume technical specifications governing the survey, planning, structural design,
and maintenance of river infrastructure---including levees, revetments, weirs, pump stations,
dams, sabo (erosion-control) dams, bed-stabilisers, landslide prevention facilities, and
steep-slope protection structures.

Practitioners---civil engineers, maintenance inspectors, and dam operators---regularly need rapid,
accurate answers to detailed technical questions about inspection intervals, design criteria,
maintenance planning procedures, and hazard-response protocols embedded in these multi-hundred-page
standards.
Retrieval from printed documents is time-consuming; large language models (LLMs) trained on
general corpora lack the domain specificity necessary for precise answers;
and commercial API-based solutions are impractical in secure government or field-deployment
environments.

\paragraph{Problem: Domain-Specific Hallucination.}
Modern LLMs are trained on vast corpora drawn from the open internet.
However, the knowledge and accumulated engineering experience of the construction domain
are only fragmentarily represented in publicly available text;
proprietary design calculations, site inspection records, and maintenance decision logs
remain largely inaccessible to general pre-training.
As a consequence, when LLMs are applied to construction design and infrastructure maintenance tasks,
their outputs do not reliably reflect domain-specific standards:
\emph{domain-specific hallucinations}---technically plausible but factually incorrect responses
that contradict the normative content of the applicable technical standards---appear frequently
and pose serious risks in safety-critical regulatory contexts.
What is needed is an LLM application \emph{grounded} in the knowledge of construction technical
standards, capable of producing accurate, evidence-traceable answers to domain questions.

\paragraph{Proposed Approach.}
This work addresses the hallucination problem through two complementary grounding strategies,
both rooted in a structured \textbf{construction domain knowledge graph} extracted from the standards:
\begin{enumerate}
  \item \textbf{GraphRAG} --- The knowledge graph, stored as a Neo4j property graph, is
        queried at inference time to inject domain-grounded context into the LLM prompt,
        suppressing hallucination by anchoring generation to graph-verified facts.
  \item \textbf{QLoRA Fine-Tuning} --- Domain QA training pairs are generated automatically
        from the knowledge graph relations and used to fine-tune a smaller LLM with QLoRA,
        internalising domain knowledge into the model weights for faster and more memory-efficient
        deployment without runtime retrieval.
\end{enumerate}
By comparing these two approaches---and a plain-LLM baseline---on a 100-question benchmark,
this paper aims to clarify \textbf{how a construction domain knowledge graph should be established
as a foundational infrastructure} for LLM-based and Agentic AI applications in the construction
and civil engineering domain.

This work investigates three strategies for building a \textbf{local, offline knowledge QA system}
over the River and Sediment Control Technical Standards using entirely open-source components:

\begin{enumerate}
  \item \textbf{Case A --- Plain LLM baseline.}
        A large general-purpose Japanese LLM (GPT-OSS Swallow~20B, Q4\_K\_M quantised) is
        queried directly with no domain-specific augmentation.

  \item \textbf{Case B --- QLoRA domain fine-tuning.}
        A smaller LLM (Swallow-8B-Instruct) is fine-tuned with QLoRA on 715 domain QA pairs
        automatically generated from a Neo4j knowledge graph constructed from the standards.
        At inference time, no external retrieval is required.

  \item \textbf{Case C --- GraphRAG.}
        The same 20B baseline model is augmented with a Neo4j knowledge graph at inference time.
        Relevant graph context (up to 2,000 characters) is retrieved via five specialised
        Cypher queries and injected into the LLM prompt.
\end{enumerate}

The three cases are evaluated on a 100-question benchmark covering all technical domains of the
standards, with automated scoring by an independent LLM judge.
Our main contributions are:

\begin{enumerate}
  \item A complete engineering methodology for building a domain-specific knowledge QA system
        over Japanese regulatory technical standards, using only open-source LLMs and local
        infrastructure.
  \item Empirical evidence that QLoRA fine-tuning of a small (8B) model on graph-derived
        training data can decisively outperform both a plain large (20B) LLM and a GraphRAG
        system using the same large LLM.
  \item A reusable schema for constructing a property knowledge graph from structured Markdown
        technical standards, covering 11 relation types and 9 node categories.
  \item Documentation of practical engineering lessons: training data generation pitfalls
        (JSON parsing errors, LLM timeout, data imbalance), Windows-specific GGUF
        quantisation workflow, and hallucination failure modes
        (see Supplementary Materials).
\end{enumerate}

Section~\ref{sec:related} reviews related work.
Section~\ref{sec:problem} defines the knowledge QA task.
Section~\ref{sec:graph} describes the knowledge graph construction.
Section~\ref{sec:method} details each of the three approaches.
Section~\ref{sec:experiments} presents the experimental setup.
Section~\ref{sec:results} reports quantitative results.
Section~\ref{sec:qualitative} conducts qualitative analysis on 10 representative questions.
Section~\ref{sec:lessons} distils high-level engineering lessons;
implementation pitfalls and toolchain details are in Supplementary Materials.
Section~\ref{sec:discussion} discusses implications and limitations.
Section~\ref{sec:conclusion} concludes.

\section{Related Work}
\label{sec:related}

\subsection{Retrieval-Augmented Generation}

Retrieval-Augmented Generation (RAG)~\cite{lewis2020rag} improves LLM factual accuracy by
retrieving relevant passages from an external corpus at inference time and conditioning
generation on the retrieved context.
Gao et al.~\cite{gao2024ragsurvey} provide a comprehensive survey of RAG architectures,
classifying retrieval strategies, augmentation methods, and generation paradigms across
diverse domains.
GraphRAG~\cite{edge2024graphrag} extends this paradigm by structuring the retrieval corpus
as a knowledge graph, enabling multi-hop relational retrieval that is not directly supported
by embedding-based nearest-neighbour search.
Jiang et al.~\cite{jiang2023active} propose Active Retrieval Augmented Generation (FLARE),
which iteratively retrieves documents based on upcoming sentence predictions, improving
coherence in long-form generation.
Self-RAG~\cite{asai2024selfrag} trains a single LLM to adaptively retrieve passages and
critique its own output through special reflection tokens, outperforming standard RAG on
diverse knowledge-intensive tasks.
Siriwardhana et al.~\cite{siriwardhana2023domainrag} demonstrate that domain-adaptive
fine-tuning of both the retriever and generator significantly improves RAG performance
on specialised corpora, a key motivation for our QLoRA fine-tuning approach.
Pan et al.~\cite{pan2024unifying} survey the intersection of LLMs and knowledge graphs,
identifying graph-augmented generation as a key research direction.
HippoRAG~\cite{guo2024hipporag} proposes neurobiologically-inspired graph-based long-term
memory for LLMs.

\subsection{Knowledge Graph Construction and Prompting}

Trajanoska et al.~\cite{trajanoska2023kg} demonstrate that LLMs can be prompted with
in-context examples to extract entities and relations from domain text, facilitating
automated knowledge graph construction from unstructured corpora.
MindMap~\cite{wen2024mindmap} converts retrieved KG triples into a graph-structured prompt
that guides the LLM through graph-of-thoughts reasoning, achieving strong results on
multi-hop medical and commonsense QA benchmarks.
Wang et al.~\cite{wang2024kgp} formalise Knowledge Graph Prompting (KGP) for
multi-document QA, showing that KG-guided retrieval outperforms dense retrieval when
documents share latent relational structure.
Besta et al.~\cite{besta2024got} introduce Graph of Thoughts (GoT), a generalisation of
chain-of-thought that represents the reasoning process as a directed acyclic graph,
enabling arbitrary aggregation and refinement of partial answers.
In our pipeline, these complementary ideas converge: we construct a domain KG from structured standards
documents, use Cypher-based multi-hop retrieval, surface KG context to the LLM as
structured triples, and evaluate reasoning quality with an independent judge.

\subsection{Parameter-Efficient Fine-Tuning}

Low-Rank Adaptation (LoRA)~\cite{hu2022lora} injects trainable low-rank matrices into frozen
pre-trained weight matrices, enabling domain adaptation with orders-of-magnitude fewer
trainable parameters than full fine-tuning.
QLoRA~\cite{dettmers2023qlora} extends LoRA to quantised (4-bit NF4) base models, enabling
fine-tuning of large models on consumer GPUs.
QA-LoRA~\cite{xu2023qalora} further co-designs the quantisation and adaptation process to
maintain model quality under aggressive bit-width reduction.
GaLore~\cite{zhao2024galore} reduces memory consumption during full-rank training via gradient
low-rank projection, achieving near-full-finetuning quality at a fraction of the memory cost.
LlamaFactory~\cite{zheng2024llamafactory} provides a unified, hardware-agnostic framework
covering over 100 model architectures and multiple PEFT strategies including LoRA, QLoRA,
and GaLore, simplifying reproducible fine-tuning experiments.
\texttt{unsloth}~\cite{han2023unsloth} provides a highly optimised QLoRA training framework
with 2--5$\times$ speedup and 70\% memory reduction relative to standard HuggingFace
\texttt{transformers} + \texttt{peft} training.
We leverage QLoRA via \texttt{unsloth} to fine-tune the 8B Swallow model on 715 domain
QA pairs with a single consumer GPU.

\subsection{Japanese LLMs}

Touvron et al.~\cite{touvron2023llama2} release Llama~2, an open-weight foundation model
that serves as the architectural template for subsequent Japanese-adapted variants.
The Swallow family~\cite{okazaki2024swallow} adapts the Llama~3 architecture~\cite{meta2024llama3}
to Japanese through continued pre-training on a large Japanese web corpus.
Fujii et al.~\cite{fujii2024continual} systematically analyse continual pre-training
strategies for cross-lingual adaptation, demonstrating that carefully scheduled vocabulary
extension and learning rate warmup are critical for retaining English capabilities while
acquiring Japanese fluency — findings directly relevant to the Swallow training recipe.
GPT-OSS Swallow-20B-RL-v0.1 is a 20B-parameter reinforcement-learning post-trained variant
released under Apache~2.0 by Tokyo Tech and AIST.
Swallow-8B-Instruct is a 8B instruction-tuned variant used as the base for QLoRA fine-tuning
in Case~B.

\subsection{LLM-as-Judge and Evaluation}

Automated evaluation of LLM outputs using a stronger judge LLM was formalised by
Zheng et al.~\cite{zheng2023judging} with the MT-Bench benchmark.
RAGAS~\cite{es2023ragas} provides a framework for automated RAG evaluation using
faithfulness, answer relevancy, and context recall metrics.

\subsection{Domain-Specific LLM Applications}

Zhao et al.~\cite{zhao2023llmsurvey} provide a broad survey of LLM capabilities and
limitations, highlighting that highly technical, low-resource domains remain challenging
and that domain-specific pre-training or fine-tuning is typically necessary for acceptable
performance.
Singhal et al.~\cite{singhal2023med} demonstrate that LLMs can encode substantial clinical
knowledge and, when augmented with retrieval and rubric-based evaluation, achieve
expert-level performance on medical licensing examinations — a paradigm closely analogous
to our river and sediment control standards QA task.
Together, these works motivate combining structured knowledge retrieval (GraphRAG) with
lightweight domain fine-tuning (QLoRA) for low-resource Japanese technical corpora.
We adopt the LLM-as-Judge approach with a rubric scored 0--3, using Qwen2.5-14B as an
independent judge model to eliminate self-scoring bias.

\section{Problem Formulation}
\label{sec:problem}

\paragraph{Task.}
Given a natural-language question $q$ about the River and Sediment Control Technical Standards
(in Japanese), generate an answer $a$ that is technically accurate, specific, and grounded in
the standards.

\paragraph{Evaluation.}
Answers are scored by an independent LLM judge $\mathcal{J}$ (Qwen2.5-14B) on a 0--3 rubric:
\begin{itemize}
  \item \textbf{3} --- Technically accurate and specific; cites standard names, chapter numbers,
        or technical concepts correctly.
  \item \textbf{2} --- Mostly correct but lacks supporting evidence or specificity.
  \item \textbf{1} --- Partially correct; contains important errors or omissions.
  \item \textbf{0} --- No answer, or technically incorrect.
\end{itemize}

\paragraph{Test set.}
The 100 test questions were generated \emph{independently} of the 715 training QA pairs.
Training data are generated automatically from KG triples via the 20B model
(Eq.~\ref{eq:data_gen}), whereas test questions are manually curated to cover
diverse question types across 8 technical categories:
Maintenance (River, Dam, Sabo), Survey, Planning, Design,
Cross-domain comparison, and Hazard.
Questions are constructed to test factual recall, procedural understanding,
multi-facility comparison, and hazard-response reasoning.
Test questions are additionally deduplicated against training data using character bigram Jaccard similarity
(threshold $\geq 0.45$ triggers exclusion) to prevent data leakage.

\paragraph{Fairness.}
The judge model (Qwen2.5-14B) is intentionally different from both the baseline and
fine-tuned inference models (Swallow series), eliminating self-scoring bias.

\subsection*{Formal Definition}

Let $\mathcal{Q}$ denote the space of natural-language questions and
$\mathcal{A}$ the space of text answers.
The knowledge QA task is to learn a mapping
\begin{equation}
  f : \mathcal{Q} \to \mathcal{A},
  \label{eq:task}
\end{equation}
such that the expected judge score is maximised:
\begin{equation}
  f^* = \arg\max_{f}\; \mathbb{E}_{q \sim \mathcal{Q}_{\mathrm{test}}}\bigl[\mathcal{J}(q,\, f(q))\bigr],
  \label{eq:objective}
\end{equation}
where the judge function $\mathcal{J}: \mathcal{Q}\times\mathcal{A}\to\{0,1,2,3\}$
assigns an ordinal quality score (rubric defined above).

The aggregate performance of a system $X$ over the 100-question test set is
\begin{equation}
  \bar{s}_X = \frac{1}{|\mathcal{Q}_{\mathrm{test}}|}
               \sum_{q\,\in\,\mathcal{Q}_{\mathrm{test}}} \mathcal{J}\bigl(q,\, f_X(q)\bigr),
  \qquad X \in \{A, B, C\}.
  \label{eq:avg_score}
\end{equation}

To prevent data leakage between training pairs $\mathcal{D}_{\mathrm{train}}$ and the test set,
pairs are deduplicated by character bigram Jaccard similarity:
\begin{equation}
  J_{\mathrm{bi}}(q_i, q_j)
  = \frac{|B(q_i)\cap B(q_j)|}{|B(q_i)\cup B(q_j)|},
  \label{eq:jaccard}
\end{equation}
where $B(q)$ is the multiset of character bigrams of $q$.
A question $q_j \in \mathcal{Q}_{\mathrm{test}}$ is excluded if
$\exists\,(q_i, a_i) \in \mathcal{D}_{\mathrm{train}}$ such that $J_{\mathrm{bi}}(q_i, q_j) \geq 0.45$.

\section{Knowledge Graph Construction}
\label{sec:graph}

\subsection{Source Documents}

The source corpus consists of eight Markdown-formatted technical standard documents
covering the River and Sediment Control Technical Standards:

\begin{itemize}
  \item Survey edition (river/dam/sabo hydrology, topography, sediment)
  \item Planning edition (river plan, sabo plan, dam plan)
  \item Design edition (levee, revetment, sabo dam, dam, landslide, steep slope)
  \item Maintenance edition (river, dam, sabo --- including inspection, life-extension,
        sedimentation management)
\end{itemize}

\subsection{Graph Formalisation}

The knowledge graph is formally defined as a typed property graph
\begin{equation}
  \mathcal{G} = (\mathcal{V},\; \mathcal{E},\; \tau_V,\; \tau_E),
  \label{eq:kg}
\end{equation}
where:
\begin{itemize}
  \item $\mathcal{V} = \mathcal{V}_S \cup \mathcal{V}_D$ is the node set,
        partitioned into structural nodes $\mathcal{V}_S$ ($|\mathcal{V}_S|=141$)
        and domain nodes $\mathcal{V}_D$ ($|\mathcal{V}_D|=59$), totalling
        $|\mathcal{V}|=200$.
  \item $\mathcal{E} \subseteq \mathcal{V}\times\mathcal{R}\times\mathcal{V}$
        is the directed edge set with $|\mathcal{E}|=268$.
  \item $\tau_V : \mathcal{V} \to \mathcal{T}_V$ assigns one of nine node types
        ($|\mathcal{T}_V|=9$).
  \item $\tau_E : \mathcal{E} \to \mathcal{R}$ assigns one of eleven relation types
        ($|\mathcal{R}|=11$).
\end{itemize}

The structural hierarchy forms a directed acyclic path graph over $\mathcal{V}_S$:
\begin{equation}
  \mathrm{Std}
  \xrightarrow{\texttt{HAS\_CHAPTER}}
  \mathrm{Ch}
  \xrightarrow{\texttt{HAS\_SECTION}}
  \mathrm{Sec}
  \xrightarrow{\texttt{HAS\_ITEM}}
  \mathrm{Item}.
  \label{eq:hierarchy}
\end{equation}
Domain nodes in $\mathcal{V}_D$ are connected to $\mathcal{V}_S$ via cross-link relations
$\texttt{DESCRIBED\_IN}$ and $\texttt{DEFINED\_IN}$, enabling
the retrieval of both semantic entities and their documental grounding in a single graph traversal.

\subsection{Node Schema}

The knowledge graph uses a property graph model in Neo4j~\cite{neo4j2024} with
\textbf{200 nodes} across nine node types:

\begin{itemize}
  \item \textbf{Structural nodes}: Standard (7), Chapter (76), Section (33), Item (25)
  \item \textbf{Domain nodes}: FacilityType (20), HazardType (8), TechnicalConcept (22),
        RequirementType (5), ProcessConcept (4)
\end{itemize}

Structural nodes capture the document hierarchy.
Domain nodes capture the semantic entities of the standards: physical structures (facilities),
natural threats (hazards), engineering methods (technical concepts), and planning processes.

\subsection{Relation Schema}

\textbf{268 relations} are defined across 11 relation types (Table~\ref{tab:relations}):

\begin{table*}[t]
  \centering
  \caption{Knowledge graph relation schema (268 relations total).}
  \label{tab:relations}
  \begin{minipage}[t]{0.48\linewidth}
    \centering
    \begin{tabular}{lll}
      \toprule
      Relation & From $\to$ To & Role \\
      \midrule
      \texttt{HAS\_CHAPTER}  & Std $\to$ Ch      & Doc.\ structure \\
      \texttt{HAS\_SECTION}  & Ch $\to$ Sec      & Doc.\ structure \\
      \texttt{HAS\_ITEM}     & Sec $\to$ Item    & Doc.\ structure \\
      \texttt{DESCRIBED\_IN} & Fac $\to$ Ch/Sec  & Rule location \\
      \texttt{SUBJECT\_TO}   & Fac $\to$ Haz     & Applicable hazard \\
      \texttt{MITIGATES}     & Fac $\to$ Haz     & Countermeasure \\
      \bottomrule
    \end{tabular}
  \end{minipage}
  \hfill
  \begin{minipage}[t]{0.48\linewidth}
    \centering
    \begin{tabular}{lll}
      \toprule
      Relation & From $\to$ To & Role \\
      \midrule
      \texttt{REQUIRES}   & Fac $\to$ Tech      & Req.\ technique \\
      \texttt{DEFINED\_IN}& Tech $\to$ Ch/Sec   & Def.\ location \\
      \texttt{USED\_IN}   & Tech $\to$ Proc     & Process stage \\
      \texttt{PRECEDES}   & Proc $\to$ Proc     & Process order \\
      \texttt{AFFECTS}    & Haz $\to$ Fac       & Impact relation \\
      \bottomrule
    \end{tabular}
  \end{minipage}
\end{table*}

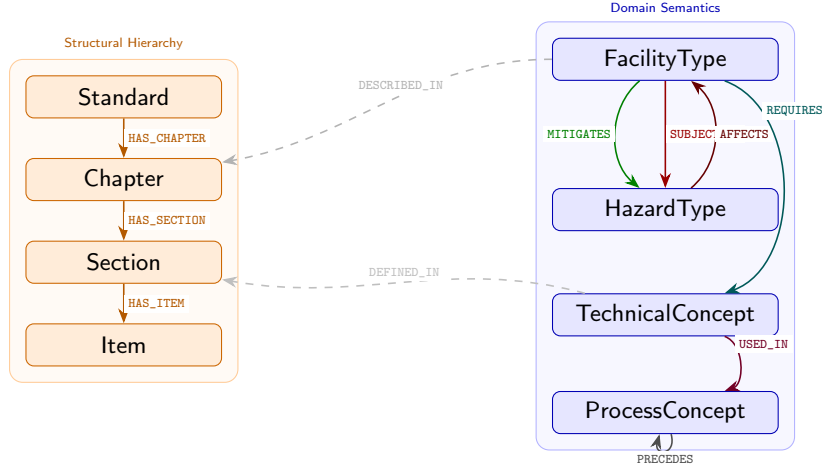
\begin{figure*}[t]
  \centering
  \begin{tikzpicture}[
    >=Stealth,
    struct/.style={rectangle, draw=orange!80!black, fill=orange!15, rounded corners=3pt,
                   minimum width=2.6cm, minimum height=0.56cm, font=\small\sffamily},
    domnd/.style={rectangle, draw=blue!70!black, fill=blue!10, rounded corners=3pt,
                   minimum width=3.0cm, minimum height=0.56cm, font=\small\sffamily},
    arr/.style={->, semithick},
    lbl/.style={font=\tiny\sffamily, fill=white, inner sep=1.5pt}
  ]
    \node[struct] (std)  at (0,  0.0) {Standard};
    \node[struct] (ch)   at (0, -1.1) {Chapter};
    \node[struct] (sec)  at (0, -2.2) {Section};
    \node[struct] (itm)  at (0, -3.3) {Item};

    \draw[arr, orange!70!black]
      (std) -- node[lbl, right]{\texttt{HAS\_CHAPTER}} (ch);
    \draw[arr, orange!70!black]
      (ch)  -- node[lbl, right]{\texttt{HAS\_SECTION}} (sec);
    \draw[arr, orange!70!black]
      (sec) -- node[lbl, right]{\texttt{HAS\_ITEM}} (itm);

    \node[domnd] (fac) at (7.2,  0.5) {FacilityType};
    \node[domnd] (haz) at (7.2, -1.5) {HazardType};
    \node[domnd] (tec) at (7.2, -2.9) {TechnicalConcept};
    \node[domnd] (prc) at (7.2, -4.2) {ProcessConcept};

    \draw[arr, green!50!black, out=220, in=140]
      (fac) to node[lbl, left]{\texttt{MITIGATES}} (haz);
    \draw[arr, red!60!black]
      (fac) -- node[lbl, right, pos=0.5]{\texttt{SUBJECT\_TO}} (haz);
    \draw[arr, red!40!black, out=40, in=320]
      (haz) to node[lbl, right]{\texttt{AFFECTS}} (fac);
    \draw[arr, teal!70!black, out=-20, in=20]
      (fac) to node[lbl, right, pos=0.2]{\texttt{REQUIRES}} (tec);
    \draw[arr, purple!60!black, out=-20, in=20]
      (tec) to node[lbl, right, pos=0.2]{\texttt{USED\_IN}} (prc);
    \draw[arr, gray!60!black, loop below, looseness=5]
      (prc) to node[lbl, below]{\texttt{PRECEDES}} (prc);

    \draw[arr, gray!60, dashed]
      (fac) to[out=180, in=10]
      node[lbl, above, pos=0.45]{\texttt{DESCRIBED\_IN}} (ch);
    \draw[arr, gray!50, dashed]
      (tec) to[out=165, in=350]
      node[lbl, above, pos=0.5]{\texttt{DEFINED\_IN}} (sec);

    \begin{scope}[on background layer]
      \node[draw=orange!40, fill=orange!4, rounded corners=5pt,
            fit=(std)(itm), inner sep=6pt,
            label={[font=\tiny\sffamily, orange!70!black]above:Structural Hierarchy}] {};
      \node[draw=blue!30, fill=blue!3, rounded corners=5pt,
            fit=(fac)(prc), inner sep=6pt,
            label={[font=\tiny\sffamily, blue!70!black]above:Domain Semantics}] {};
    \end{scope}
  \end{tikzpicture}
  \caption{Knowledge graph schema (Node \& Relation Map).
    The \emph{Structural Hierarchy} (left, orange) encodes four-level document
    structure. The \emph{Domain Semantics} (right, blue) encodes engineering entities
    and their mutual relations. Dashed grey arrows cross-link domain nodes to
    structural locations (\texttt{DESCRIBED\_IN}, \texttt{DEFINED\_IN}).
    In total: 9 node types, 200 nodes, 11 relation types, 268 relations.}
  \label{fig:schema}
\end{figure*}

Figure~\ref{fig:schema} illustrates the complete node-and-relation schema.
As shown in the figure, structural nodes form a four-level document hierarchy
(Standard $\to$ Chapter $\to$ Section $\to$ Item), while domain nodes capture the
engineering semantics (facilities, hazards, concepts, and processes) with
cross-links into the structural hierarchy via \texttt{DESCRIBED\_IN} and
\texttt{DEFINED\_IN}.

The graph is loaded from four hand-curated CSV files
(\texttt{nodes\_standard.csv}, \texttt{nodes\_chapter\_section\_item.csv},
\texttt{nodes\_domain.csv}, \texttt{relations.csv}) via the script
\texttt{02\_load\_neo4j.py}, which applies \texttt{MERGE} deduplication and
enforces B-tree and fulltext indexes at schema initialisation.

\section{Methodology}
\label{sec:method}

\subsection{End-to-End Algorithmic Formulation}
\label{sec:formulation}

All three experimental cases share the same input $q\in\mathcal{Q}$ and the same
judge evaluation $\mathcal{J}$ (Eq.~\ref{eq:avg_score}), but differ in how $f(q)$ is computed.

\paragraph{Case A --- Plain LLM (baseline).}
A frozen 20B-parameter model $\mathrm{LLM}_{20B}$ with weights $\theta_{20B}$ generates
the answer directly:
\begin{equation}
  a_A = \mathrm{LLM}_{20B}\!\left(p_{\mathrm{sys}} \oplus q;\; \theta_{20B}\right),
  \label{eq:case_a}
\end{equation}
where $\oplus$ denotes prompt concatenation and $p_{\mathrm{sys}}$ is a fixed system prompt.

\paragraph{Case C --- GraphRAG.}
At inference time, the graph $\mathcal{G}$ is queried to retrieve domain context.
Let $\mathcal{K}=\mathrm{ExtractKeywords}(q)$ denote the keyword set extracted from $q$.
Five typed Cypher queries are executed in parallel:
\begin{equation}
  \mathcal{R}_i = \mathrm{Query}_i(\mathcal{K},\mathcal{G}),\quad i=1,\ldots,5.
  \label{eq:queries}
\end{equation}
Retrieved records are merged and de-duplicated:
$\mathcal{R} = \mathrm{Dedup}\!\left(\bigcup_{i=1}^{5}\mathcal{R}_i\right)$.
Each record $r\in\mathcal{R}$ receives a score
\begin{equation}
  \sigma(r) =
  \begin{cases}
    s_{\mathrm{neo4j}}(r)\times 10 & r \in \mathcal{R}_1 \;\text{(fulltext)}, \\
    \mathrm{KeywordMatch}(r,\mathcal{K}) & \text{otherwise}.
  \end{cases}
  \label{eq:score}
\end{equation}
If $|\mathcal{R}|<25$, an adaptive retry is triggered with doubled retrieval depth $k'=2k$
and broad substring matching, updating $\mathcal{R}$ before scoring.
The ranked context is then constructed as
\begin{multline}
  \mathrm{ctx} = \mathrm{Truncate}\!\Bigl(
    \mathrm{Concat}\bigl(\mathrm{Top}_{80\%}\{r: \sigma(r)>0\}\bigr),\\
    L_{\max}=2{,}000\text{ chars}\Bigr),
  \label{eq:ctx}
\end{multline}
and the answer is generated as
\begin{equation}
  a_C = \mathrm{LLM}_{20B}\!\left(\mathrm{ctx} \oplus p_{\mathrm{sys}} \oplus q;\; \theta_{20B}\right).
  \label{eq:case_c}
\end{equation}

\paragraph{Case B --- QLoRA Fine-Tuning.}
Training data $\mathcal{D}_{\mathrm{train}}$ is generated from the edge set
$\mathcal{T}=\{(s,r,t)\in\mathcal{E}\}$ ($|\mathcal{T}|=268$).
For each triple, the 20B model generates candidate QA pairs:
\begin{equation}
  D_r = \mathrm{LLM}_{20B}^{\mathrm{gen}}\!\bigl(\mathrm{TriplePrompt}(s,r,t)\bigr)
       = \{(q_{r,j},a_{r,j})\}_{j=1}^{3}.
  \label{eq:data_gen}
\end{equation}
After bigram-Jaccard deduplication (Eq.~\ref{eq:jaccard}, threshold $<0.45$):
\begin{equation}
  \mathcal{D}_{\mathrm{train}}
  = \mathrm{Dedup}\!\Bigl(\bigcup_{r\in\mathcal{T}} D_r\Bigr),\quad
  |\mathcal{D}_{\mathrm{train}}| = 715.
  \label{eq:dedup}
\end{equation}
LoRA adapters $\Delta W = BA$ ($B\in\mathbb{R}^{d\times\rho}$, $A\in\mathbb{R}^{\rho\times d}$,
rank $\rho\ll d$) are injected into the frozen 8B base weights $\theta_{\mathrm{base}}^{8B}$
and optimised by minimising the causal language-modelling loss:
\begin{equation}
  \mathcal{L}(\Delta W) =
  -\sum_{(q,a)\,\in\,\mathcal{D}_{\mathrm{train}}}
  \log P\!\bigl(a\mid q;\; \theta_{\mathrm{base}}^{8B} + \Delta W\bigr).
  \label{eq:lora_loss}
\end{equation}
At inference, the fine-tuned answer is
\begin{equation}
  a_B = \mathrm{LLM}_{8B}\!\left(q;\; \theta_{\mathrm{base}}^{8B} + \Delta W^*\right),
  \label{eq:case_b}
\end{equation}
where $\Delta W^* = \arg\min_{\Delta W}\mathcal{L}(\Delta W)$.
Equations~\ref{eq:case_a}--\ref{eq:case_b} together define the complete
input-to-output pipeline for the three cases evaluated in this paper.

\subsection{Case A --- Plain LLM Baseline}
\label{sec:case_a}

Case~A serves as the baseline.
The GPT-OSS Swallow-20B-RL-v0.1 model (Q4\_K\_M quantisation, 15.8~GB) is served locally
via Ollama and queried directly with the user's question.

A critical implementation detail of GPT-OSS models is that they use a proprietary channel-based
chat template (\texttt{<|channel|>final}) that is not supported by Ollama's OpenAI-compatible
\texttt{/v1/chat/completions} endpoint (which returns empty content).
We bypass this by calling Ollama's native \texttt{/api/generate} endpoint with
\texttt{raw=True} and a manually constructed prompt template:
\begin{verbatim}
<|start|>system<|message|>{system}<|end|>
<|start|>user<|message|>{user}<|end|>
<|start|>assistant<|channel|>final<|message|>
\end{verbatim}

Generation parameters: temperature 0.2, repeat penalty 1.2, max context 8,192 tokens.
The repeat penalty proved critical---without it, the model exhibited runaway repetition
on domain-specific prompts (e.g., repeating the phrase ``long-life extension'' 300$+$ times in Japanese, yielding score~0).

\subsection{Case C --- GraphRAG}
\label{sec:case_c}

Case~C augments Case~A with a Neo4j-based retrieval pipeline.
At inference time, five specialised Cypher queries are executed in parallel
to gather domain-specific context:

\begin{enumerate}
  \item \textbf{Fulltext keyword search} --- Lucene full-text index over node names/descriptions.
  \item \textbf{Facility context} --- 2-hop subgraph: FacilityType $\to$ Chapter/Section $\to$ TechnicalConcept.
  \item \textbf{Hazard--facility map} --- HazardType $\to$ FacilityType $\to$ Chapter.
  \item \textbf{Maintenance cycle query} --- ProcessConcept chain via \texttt{PRECEDES}.
  \item \textbf{Facility comparison} --- Multi-facility cross-referencing.
\end{enumerate}

Results are deduplicated and re-ranked by a scoring function:
fulltext hits receive $\text{Neo4j score} \times 10$;
other queries receive the keyword match count.
Records with score $> 0$ are retained; the top 80\% by score are selected.
The resulting context is truncated to 2,000 characters and prepended to the Case~A prompt.

\paragraph{Adaptive retry.}
If the number of graph hits falls below 25,
the pipeline retries with $\text{TOP\_K} \times 2$ and a broad section search
(substring match on Chapter, Section, TechnicalConcept node names)
to improve recall on under-specified queries.
Figure~\ref{fig:graphrag_flow} illustrates the complete inference-time retrieval pipeline for Case~C.
Detailed discussion of the design decisions is provided in Supplementary~S4.

\begin{figure*}[t]
  \centering
  \begin{tikzpicture}[
    >=Stealth, node distance=0.70cm,
    terminal/.style={rectangle, draw=gray!70, fill=gray!12, rounded corners=4pt,
                     minimum width=4.8cm, minimum height=0.58cm, font=\small\sffamily},
    process/.style={rectangle, draw=blue!60!black, fill=blue!9, rounded corners=2pt,
                     minimum width=4.8cm, minimum height=0.58cm, font=\small\sffamily},
    decision/.style={diamond, draw=red!60!black, fill=red!9, aspect=2.8,
                     font=\small\sffamily},
    retry/.style={rectangle, draw=orange!70!black, fill=orange!10, rounded corners=3pt,
                  minimum width=4.4cm, minimum height=0.58cm, font=\small\sffamily, align=center},
    llmbox/.style={rectangle, draw=purple!60!black, fill=purple!8, rounded corners=2pt,
                   minimum width=4.8cm, minimum height=0.58cm, font=\small\sffamily},
    judgebox/.style={rectangle, draw=green!50!black, fill=green!7, rounded corners=2pt,
                     minimum width=4.8cm, minimum height=0.58cm, font=\small\sffamily},
    arr/.style={->, semithick},
    lbl/.style={font=\tiny\sffamily, fill=white, inner sep=1pt}
  ]
    \node[terminal]                       (q)     {User Question};
    \node[process,   below=of q]          (kw)    {\texttt{extract\_keywords()}};
    \node[process,   below=of kw,
          text width=4.4cm, align=center] (qry)   {5 Parallel Cypher Queries \\
          {\tiny Fulltext / Facility / Hazard / Maintenance / Comparison}};
    \node[process,   below=of qry]        (dedup) {Deduplicate \& Score};
    \node[decision,  below=of dedup]      (cond)  {hits $\geq 25$?};
    \node[process,   below=of cond]       (rank)  {Re-rank (top 80\%)};
    \node[process,   below=of rank]       (ctx)   {Build Context ($\leq$2,000 chars)};
    \node[process,   below=of ctx]        (prep)  {Prepend Context to Prompt};
    \node[llmbox,    below=of prep]       (llm)   {Swallow-20B via Ollama};
    \node[judgebox,  below=of llm]        (judge) {GPT-4o Judge ($s_{\mathrm{J}}$)};
    \node[terminal,  below=of judge]      (out)   {Score \& Final Answer};

    \node[retry, right=2.2cm of cond]    (retry)
      {Adaptive Retry \\ {\tiny TOP\_K ${\times}$2 + broad section search}};

    \draw[arr] (q)     -- (kw);
    \draw[arr] (kw)    -- (qry);
    \draw[arr] (qry)   -- (dedup);
    \draw[arr] (dedup) -- (cond);
    \draw[arr] (cond)  -- node[lbl, left]{Yes}  (rank);
    \draw[arr] (rank)  -- (ctx);
    \draw[arr] (ctx)   -- (prep);
    \draw[arr] (prep)  -- (llm);
    \draw[arr] (llm)   -- (judge);
    \draw[arr] (judge) -- (out);

    \draw[arr] (cond.east)   -- node[lbl, above]{No} (retry.west);
    \draw[arr] (retry.north) |- (dedup.east);
  \end{tikzpicture}
  \caption{GraphRAG inference pipeline (Case~C).
    At query time, keywords extracted from the user question drive five parallel
    Neo4j Cypher queries whose results are deduplicated and scored.
    If fewer than 25 hits are returned, an adaptive retry doubles TOP\_K and
    broadens the match to substring search.
    The top-scoring 80\% of records (up to 2,000 chars) are prepended to the
    plain-LLM prompt before generation by Swallow-20B.
    Qwen2.5-14B then evaluates the output and returns $s_{\mathrm{J}} \in \{0,1,2,3\}$.}
  \label{fig:graphrag_flow}
\end{figure*}

\subsection{Case B --- QLoRA Fine-Tuning}
\label{sec:case_b}

\subsubsection{Training Data Generation}
\label{sec:data_gen}

QA training pairs are generated by prompting the Swallow-20B model via the GraphRAG API
with each of the 268 graph relations as context.
For each relation triple $(s, r, t)$ (e.g., FacilityType ``Sabo-Dam'' \texttt{REQUIRES}
TechnicalConcept ``Stability-Calculation''), the model is asked to generate three QA pairs
in JSON array format.
After deduplication using character bigram Jaccard similarity (threshold $< 0.45$),
\textbf{715 unique QA pairs} are retained from a target of
$268 \times 3 = 804$.

Two main failure modes caused the 89-question shortfall:
\begin{itemize}
  \item \textbf{JSON parse errors (Extra data).}
        The model occasionally generated text after the closing bracket of the JSON array
        (e.g., references like \texttt{[River-Law Art.X]}), causing \texttt{json.loads()} to fail.
        The greedy regex \texttt{re.search(r"\textbackslash[.+\textbackslash]", ..., DOTALL)}
        captured beyond the intended boundary.
        Fix: replace with \texttt{json.JSONDecoder().raw\_decode()} from each \texttt{[} position.
  \item \textbf{Read timeouts (primary cause, $\approx$15 relations).}
        Setting \texttt{num\_predict=-1} (unlimited) caused the model to occasionally
        generate excessively long responses ($> 120$~s).
        All three retry attempts failing leads to skipping the entire relation ($\le 3$~Q lost).
        Fix: set \texttt{num\_predict=2048} and reduce to 1,024 on retry.
\end{itemize}

The per-relation-type distribution of the 715 QA pairs is shown in Table~\ref{tab:qa_dist}.

\begin{table}[H]
  \centering
  \caption{Training QA pair distribution by relation type.}
  \label{tab:qa_dist}
  \begin{tabular}{lrr}
    \toprule
    Relation Type & Count & Share (\%) \\
    \midrule
    \texttt{HAS\_CHAPTER}   & 212 & 29.7 \\
    \texttt{DESCRIBED\_IN}  &  89 & 12.4 \\
    \texttt{HAS\_SECTION}   &  86 & 12.0 \\
    \texttt{HAS\_ITEM}      &  63 &  8.8 \\
    \texttt{REQUIRES}       &  56 &  7.8 \\
    \texttt{SUBJECT\_TO}    &  55 &  7.7 \\
    \texttt{DEFINED\_IN}    &  54 &  7.6 \\
    \texttt{USED\_IN}       &  44 &  6.2 \\
    \texttt{MITIGATES}      &  28 &  3.9 \\
    \texttt{AFFECTS}        &  21 &  2.9 \\
    \texttt{PRECEDES}       &   7 &  1.0 \\
    \midrule
    Total & 715 & 100.0 \\
    \bottomrule
  \end{tabular}
\end{table}

\texttt{HAS\_CHAPTER} dominates (29.7\%) while \texttt{PRECEDES} is underrepresented (1.0\%).
Future work should up-sample rare relation types or use weighted sampling at training time.

\subsubsection{QLoRA Training Setup}

\begin{sloppypar}
The base model is \path{tokyotech-llm/Llama-3-Swallow-8B-Instruct-v0.1} loaded in
4-bit NF4 quantisation via \texttt{bitsandbytes=0.49.2}.
\end{sloppypar}
LoRA adapters are applied to all seven projection layers
(q, k, v, o, gate, up, down proj).
Hyperparameters are summarised in Table~\ref{tab:lora_hp}.

\begin{table*}[t]
  \centering
  \caption{QLoRA hyperparameters (Case B).}
  \label{tab:lora_hp}
  \begin{minipage}[t]{0.48\linewidth}
    \centering
    \begin{tabular}{ll}
      \toprule
      Parameter & Value \\
      \midrule
      \texttt{lora\_r}                      & 16 \\
      \texttt{lora\_alpha}                  & 16 \\
      \texttt{lora\_dropout}                & 0.0 \\
      Target modules                        & q/k/v/o/gate/up/down proj \\
      \texttt{load\_in\_4bit}               & True (NF4) \\
      \texttt{max\_seq\_length}             & 2,048 \\
      \texttt{num\_train\_epochs}           & 3 \\
      Batch size per device                 & 2 \\
      \bottomrule
    \end{tabular}
  \end{minipage}
  \hfill
  \begin{minipage}[t]{0.48\linewidth}
    \centering
    \begin{tabular}{ll}
      \toprule
      Parameter & Value \\
      \midrule
      Gradient accum.\ steps                & 4 (eff.\ batch $= 8$) \\
      \texttt{learning\_rate}               & 2e-4 \\
      LR scheduler                          & cosine \\
      \texttt{warmup\_ratio}                & 0.05 \\
      \texttt{weight\_decay}                & 0.01 \\
      \texttt{packing}                      & False \textsuperscript{$\dagger$} \\
      \texttt{bf16}                         & True \\
      Framework                             & \texttt{unsloth} 2026.2.1 \\
      \bottomrule
    \end{tabular}
  \end{minipage}
  \par\smallskip
  \raggedright\footnotesize $\dagger$ Disabled to avoid triton JIT hang on Windows.
\end{table*}

Training uses the Llama-3 Instruct chat format with a domain-specific Japanese system prompt
(``You are an expert well-versed in the River and Sediment Control Technical Standards
[Survey / Planning / Design / Maintenance]. Please provide accurate and practical answers.'').

\paragraph{Convergence check (4-stage subset experiment).}
To confirm convergence before committing to the full dataset,
we train on four progressively larger subsets (stratified by relation type, seed 42).
Results in Table~\ref{tab:convergence} show monotonically decreasing loss across all stages.

\begin{table}[H]
  \centering
  \caption{QLoRA convergence check on stratified subsets.}
  \label{tab:convergence}
  \begin{tabular}{rrrr}
    \toprule
    Subset & QA Pairs & Final Loss & Time \\
    \midrule
    100 & 100 & 0.7958 & 4.4 min \\
    250 & 250 & 0.6859 & 10.2 min \\
    500 & 500 & 0.6045 & 20.0 min \\
    715 & 715 & \textbf{0.5565} & 28.5 min \\
    \bottomrule
  \end{tabular}
\end{table}

\subsubsection{GGUF Quantisation and Ollama Deployment}

The LoRA adapter is merged into the full 16-bit model via \texttt{unsloth}'s
\texttt{merge\_and\_unload()} (output: $\approx$15.3~GB safetensors).
The merged model is then quantised to GGUF Q4\_K\_M ($\approx$4.7~GB) using the
official \texttt{llama.cpp} Windows binary.

\noindent\textbf{Windows-specific pitfall.}
\texttt{unsloth}'s built-in \texttt{save\_pretrained\_gguf()} attempts to install
\texttt{llama.cpp} via \texttt{apt} (Linux package manager) and hangs silently on Windows.
Workaround: use the pre-built \texttt{llama-quantize.exe} from the \texttt{llama.cpp}
Windows release directly.
Additionally, \texttt{ollama create} rejects paths under project or OneDrive directories
due to ``untrusted mount point'' restrictions;
the GGUF file must be copied to a top-level path (e.g., \texttt{C:\textbackslash{}ollama\_import\textbackslash}).

At deployment, the model is served as \texttt{swallow8b-lora-n715} via Ollama (4.9~GB).
The Modelfile embeds the domain system prompt and generation parameters:
temperature 0.3, repeat penalty 1.1.
Inference uses the same FastAPI endpoint as Case~A (\texttt{POST /query/plain})
with the model switched at configuration time.

\section{Experimental Setup}
\label{sec:experiments}

\paragraph{Hardware.}
All experiments run on a single workstation with an NVIDIA GeForce RTX~4060~Ti
(16~GB VRAM), Windows~11, Python~3.12.
LLM serving uses Ollama (CPU+GPU, auto-offload).
Neo4j~2026.01.4 Desktop is used for the knowledge graph.

\paragraph{LLM models.}
\begin{sloppypar}
\begin{itemize}
  \item Case~A-Plain / C-GraphRAG: \path{hf.co/mmnga-o/GPT-OSS-Swallow-20B-RL-v0.1-gguf:Q4_K_M}
        (15.8~GB, Apache~2.0)
  \item Case~B-QLoRA: \texttt{swallow8b-lora-n715} (4.9~GB), QLoRA FT on 715 domain QA pairs,
        base: \path{tokyotech-llm/Llama-3-Swallow-8B-Instruct-v0.1}
  \item Judge: \texttt{qwen2.5:14b} (independent model; eliminates self-scoring bias)
\end{itemize}
\end{sloppypar}

\paragraph{Evaluation procedure.}
The 100 test questions (manually curated, generated independently of the 715 training QA pairs;
see Section~\ref{sec:problem}) are posed to each case via the
FastAPI endpoint (\texttt{http://localhost:8080}).
Each answer is then submitted to the judge LLM with a scoring rubric.
Case~A and Case~C were evaluated in a single run (100~Q, JSONL streamed).
Case~B was evaluated in a separate run (\texttt{--case-b} flag, same 100~Q).

\paragraph{GraphRAG parameters.}
\texttt{GRAPH\_TOP\_K=20} (Neo4j search width per sub-query);
\texttt{GRAPH\_RERANK\_RATIO=0.8};
context cap 2,000 characters.

\section{Results}
\label{sec:results}

\subsection{Overall Scores}

Table~\ref{tab:overall} summarises the three-way comparison across all 100 questions.

\begin{table*}[t]
  \centering
  \setlength{\tabcolsep}{4pt}
  \caption{Overall evaluation results (100-question benchmark). $\Delta$ Avg = improvement over Case~A (plain LLM baseline). Low-qual.\ = score-0 or score-1 responses.}
  \label{tab:overall}
  \begin{tabular}{@{}llccccccc@{}}
    \toprule
    Case & Model & Avg & $\Delta$ Avg & Sc-3 (\%) & Sc-2 (\%) & Low-qual.\ (\%) & Lat.\ (s) & Speedup \\
    \midrule
    A (Plain LLM)         & Swallow-20B         & 2.29          & ---              & 60          & 12          & 28          & 42.2          & $1.0\times$ \\
    C (GraphRAG)          & Swallow-20B         & 2.62          & $+0.33$          & 77          & 10          & 13          & 31.1          & $1.4\times$ \\
    \textbf{B (QLoRA FT)} & \textbf{Swallow-8B} & \textbf{2.92} & $\mathbf{+0.63}$ & \textbf{92} & \textbf{8}  & \textbf{0}  & \textbf{14.2} & $\mathbf{3.0\times}$ \\
    \bottomrule
  \end{tabular}
\end{table*}

\textbf{Answer quality.}
Case~B (QLoRA FT) achieves an average score of 2.92, outperforming
Case~A by $\Delta=+0.63$ and Case~C by $\Delta=+0.30$.
The Score-3 rate rises from 60\% (A) to 77\% (C) to \textbf{92\%} (B),
while Score-2 responses remain at 8\% for Case~B with no score-0 or score-1---
a zero low-quality rate versus 28\% for Case~A and 13\% for Case~C.
The improvement from A to C (+17 pp Score-3) confirms that graph-grounded context
reduces hallucination; the further jump from C to B (+15 pp) shows that
domain fine-tuning additionally internalises technical vocabulary and procedural structure
in a way that retrieval alone cannot achieve.

\textbf{Inference efficiency.}
Case~B is \textbf{3.0$\times$ faster} than Case~A (14.2~s vs.\ 42.2~s mean latency) and
2.2$\times$ faster than Case~C (31.1~s), despite using a smaller 8B model.
Shorter, domain-tailored answers (avg 284 chars for B vs.\ 2,349 chars for C) are the
primary driver: the fine-tuned model generates concise, high-precision answers
rather than long, hedged responses.
Case~C is faster than Case~A in 96/100 questions (mean $-11.1$~s): graph context
constrains the LLM output space and reduces token generation time,
more than offsetting the Neo4j query overhead.

\textbf{Key finding.} The combination of a smaller fine-tuned model (8B) and
domain-adapted outputs dominates both accuracy \emph{and} speed---
reinforcing that QLoRA fine-tuning is a practically superior strategy
compared to larger plain or retrieval-augmented models in this domain.

\begin{figure}[H]
  \centering
  \includegraphics[width=\columnwidth]{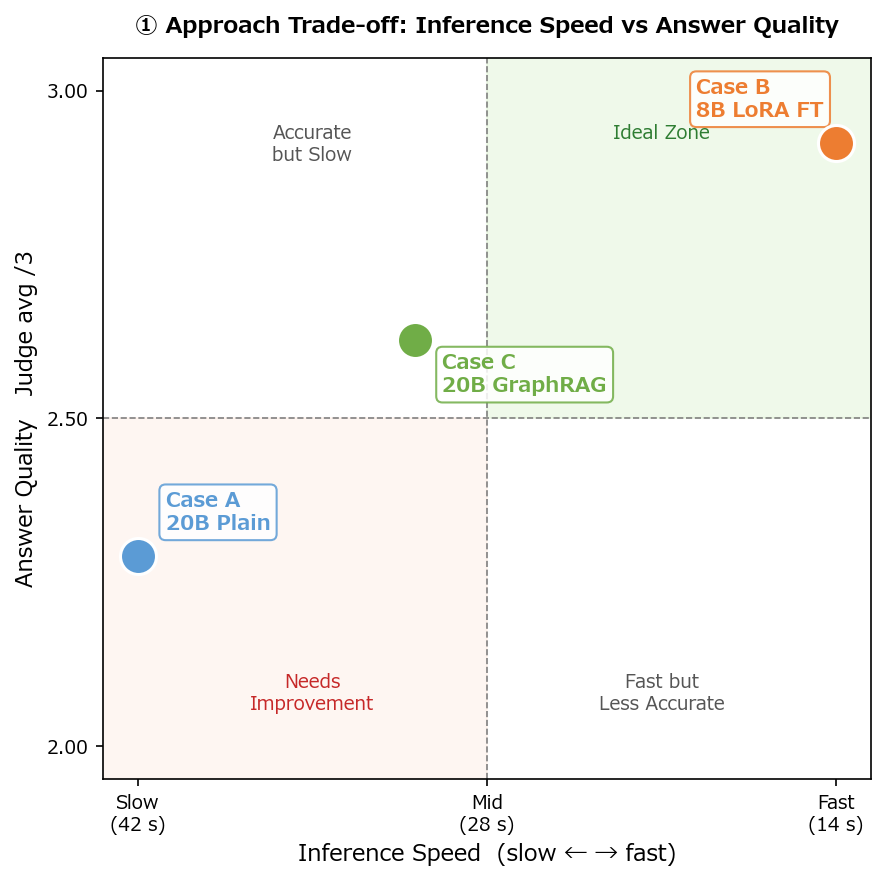}
  \caption{Approach trade-off: inference speed vs.\ answer quality (normalised).
           Case~B (QLoRA FT) occupies the ideal upper-right quadrant --- highest quality
           and fastest inference.}
  \label{fig:quadrant}
\end{figure}

\subsection{Score Distributions}

\begin{figure*}[t]
  \centering
  \includegraphics[width=\textwidth]{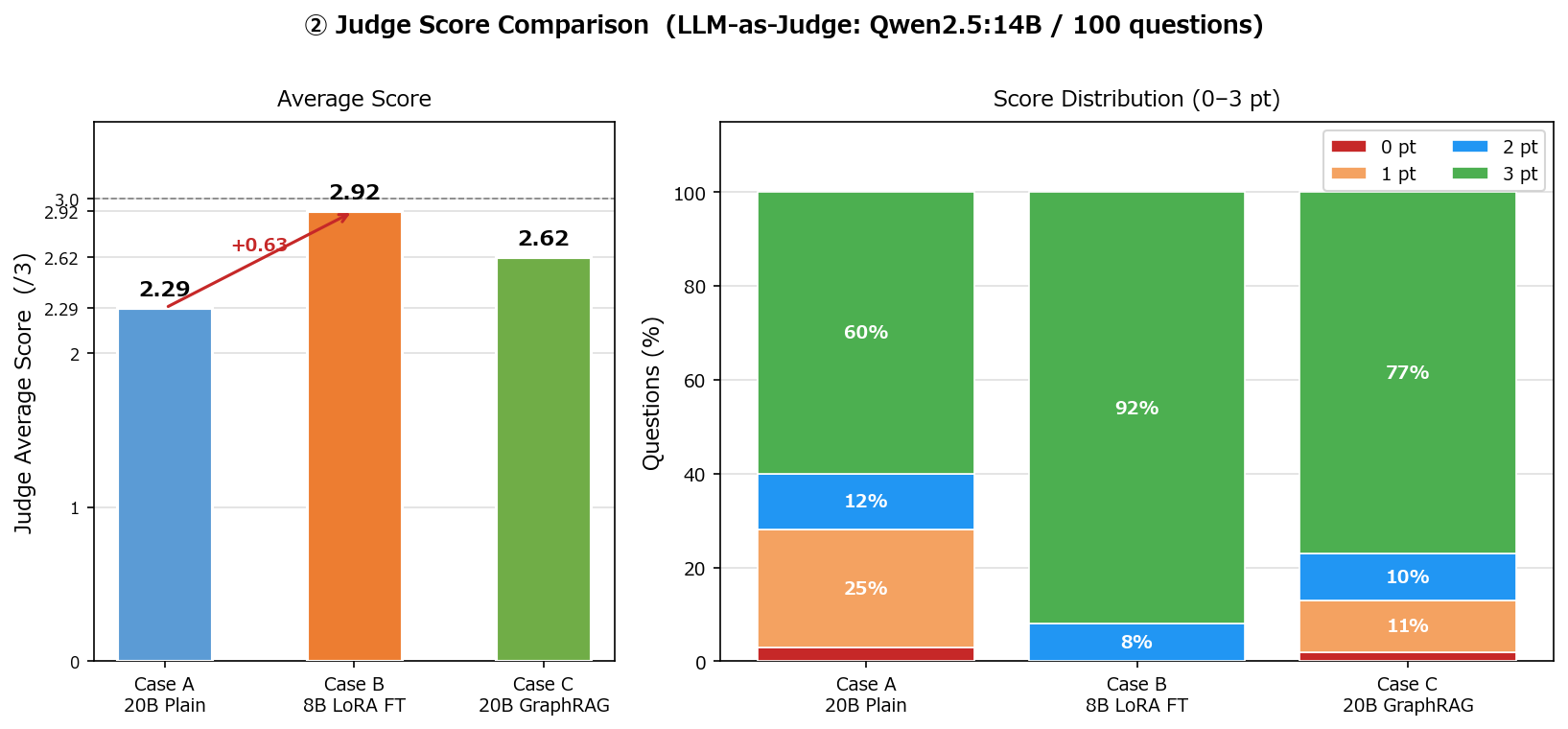}
  \caption{Judge score distributions (0--3) for Cases A, B, and C across 100 questions.}
  \label{fig:scores}
\end{figure*}

Figure~\ref{fig:scores} shows the full score distributions for all three cases.
The distributions reveal qualitatively different failure modes across the three approaches,
not merely quantitative differences in mean score.

\paragraph{Case A (Plain LLM).}
Case~A achieves 60\% score-3, but 28\% of responses are low-quality (score $\leq 1$):
3 score-0 responses result from token-repetition collapse (see Supplementary~S3),
while 25 score-1 responses are fluent but technically imprecise --- answers that describe
the correct procedure in general terms yet omit the specific standard values, chapter
references, or exception clauses required for full credit.
The distribution is notably \emph{bimodal}: the model either knows an answer well enough
to score 3, or produces a generic response that fails on technical specificity.

\paragraph{Case C (GraphRAG).}
GraphRAG retrieval reduces low-quality responses from 28\% to 13\% and raises score-3
from 60\% to 77\%. The improvement is concentrated in score-1 $\to$ score-3 uplift:
graph-grounded context supplies the specific numerical thresholds and regulatory identifiers
that were absent in plain-LLM generation.
The 13\% residual low-quality rate is dominated by score-1 responses in which the model
retrieves a relevant but slightly off-topic subgraph node and generates a partially
correct answer; score-0 drops to just 2 responses.

\paragraph{Case B (QLoRA FT).}
Case~B achieves 92 score-3 and 8 score-2 responses, with \textbf{no score-0 or score-1}.
The distribution is effectively \emph{unimodal} at score-3, a qualitative shift not
observed in the other two cases.
The complete elimination of low-quality responses indicates that QLoRA fine-tuning
internalises domain vocabulary and regulatory structure at the weight level, making
hallucination suppression independent of retrieval quality.
The 8 score-2 answers involve borderline judgment questions --- cases where multiple valid
interpretations exist within the training corpus --- rather than factual errors.

\paragraph{Key takeaway.}
The score distributions demonstrate that retrieval augmentation (Case~C) and
parameter-level domain adaptation (Case~B) address complementary failure modes.
GraphRAG is most effective at converting score-1 inadequate answers into score-3 by
providing missing factual grounding; QLoRA fine-tuning additionally eliminates
the bimodal tail by stabilising generation even on unfamiliar phrasings.

\begin{figure*}[t]
  \centering
  \includegraphics[width=\textwidth]{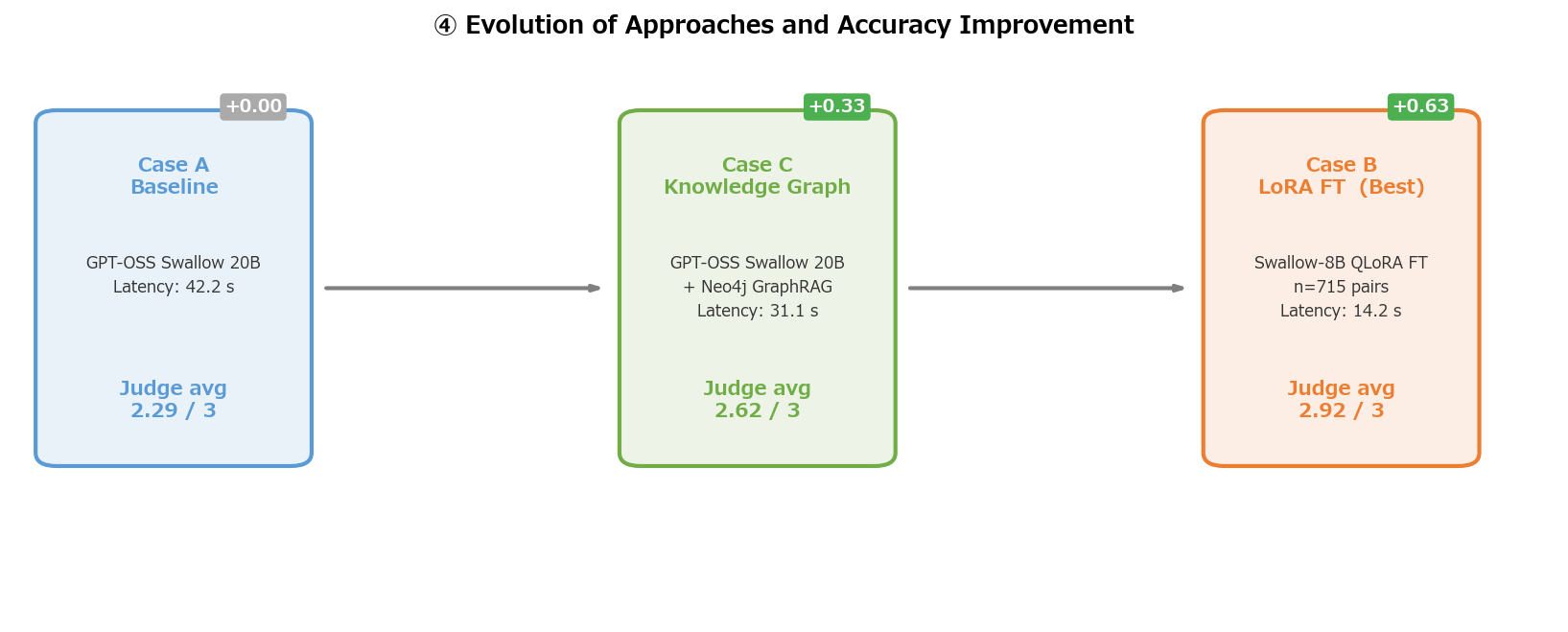}
  \caption{Evolution of approaches: accuracy improvement across experimental phases.
           QLoRA FT (Case~B) achieves the highest score at the final stage.}
  \label{fig:evolution}
\end{figure*}

\subsection{By-Category Analysis}

Table~\ref{tab:category} presents the by-category breakdown.

\begin{table}[H]
  \centering
  \caption{Judge average score (/3) by technical category.}
  \label{tab:category}
  \begin{tabular}{lcccc}
    \toprule
    Category & $N$ & A avg & B avg & C avg \\
    \midrule
    Survey              & 7  & 2.14 & \textbf{2.86} & 2.57 \\
    Planning            & 8  & 2.75 & \textbf{2.88} & 2.62 \\
    Design              & 15 & 2.13 & \textbf{2.93} & 2.60 \\
    Maint.\ (River)    & 20 & 2.05 & \textbf{2.95} & 2.55 \\
    Maint.\ (Dam)      & 15 & 2.47 & \textbf{2.93} & 2.73 \\
    Maint.\ (Sabo)     & 15 & 2.33 & \textbf{2.93} & 2.53 \\
    Hazard              & 10 & 2.40 & \textbf{2.80} & 2.60 \\
    Cross-domain        & 10 & 2.30 & \textbf{3.00} & 2.80 \\
    \midrule
    Overall             & 100 & 2.29 & \textbf{2.92} & 2.62 \\
    \bottomrule
  \end{tabular}
\end{table}

Case~B achieves the highest score in all 8 categories.
Case~B scores \textbf{3.00/3} in Cross-domain (10 Q), 2.95 in Maintenance~River (20 Q),
and 2.93 in Design and Maintenance~Dam/Sabo (15 Q each).
Case~C outperforms Case~A in all categories except Planning
(C: 2.62 vs.\ A: 2.75, $-0.13$) --- attributed to over-retrieval of
Chapter metadata nodes that misled generation.

\subsection{Latency Analysis}

Table~\ref{tab:latency} shows the latency breakdown.

\begin{table}[H]
  \centering
  \caption{Inference latency comparison.}
  \label{tab:latency}
  \begin{tabular}{lrrr}
    \toprule
    Metric (s) & Case A & Case B & Case C \\
    \midrule
    Mean    & 42.2 & \textbf{14.2} & 31.1 \\
    Median  & 43.5 & --- & 33.4 \\
    Min     & 22.0 & --- &  7.9 \\
    Max     & 43.9 & --- & 35.5 \\
    \midrule
    Total (100 Q, min) & 70.3 & \textbf{23.7} & 51.9 \\
    \bottomrule
  \end{tabular}
\end{table}

Case~C is faster than Case~A in 96/100 questions (mean $-11.1$~s):
graph context constrains the LLM output space and reduces token generation time,
outweighing the overhead of Neo4j query execution.
Case~B is fastest in all questions; shorter domain-tailored answers (avg 284 chars vs.\
2,349 chars for A and 2,452 chars for C) are the primary driver.

\subsection{Evolution of Approaches}

Figure~\ref{fig:evolution} traces the experimental trajectory from the initial 14-question
prototype to the final 100-question three-way benchmark.
A phase-by-phase breakdown is provided in Supplementary~S5.

\section{Qualitative Analysis}
\label{sec:qualitative}

We select 10 representative questions from the benchmark to contrast the three approaches
qualitatively.
Questions are selected to cover diverse score patterns:
B beats A by 3 points (Q5, Q24, Q82), A beats B (Q26, Q52, Q95),
C completely fails while B succeeds (Q37), and competitive scenarios (Q14, Q69, Q91).
Full answer texts are available in the supplementary companion document.

Table~\ref{tab:qualitative} summarises the 10 selected questions.

\begin{table}[H]
  \centering
  \caption{Scores for 10 representative questions (A/B/C, 0--3).}
  \label{tab:qualitative}
  \begin{tabular}{clccc}
    \toprule
    Q\# & Sub-category & A & B & C \\
    \midrule
    Q5  & Levee life-extension       & 0 & \textbf{3} & 3 \\
    Q14 & Cycle-type maint.\ flow   & 3 & \textbf{3} & 3 \\
    Q24 & Dam life-extension        & 0 & \textbf{3} & 3 \\
    Q26 & Dam sedimentation ratio   & \textbf{3} & 2 & 2 \\
    Q37 & Sabo dam emergency insp.  & 1 & \textbf{3} & 0 \\
    Q52 & Basin avg.\ rainfall      & \textbf{3} & 2 & 1 \\
    Q69 & Sabo dam stability calc.  & 1 & \textbf{3} & 3 \\
    Q82 & Cross-domain comparison   & 0 & \textbf{3} & 3 \\
    Q91 & Flood hazard levee        & 1 & \textbf{3} & 3 \\
    Q95 & Landslide activity class. & \textbf{3} & 2 & 1 \\
    \bottomrule
  \end{tabular}
\end{table}

\paragraph{Finding 1: Catastrophic failure of Case A on open-ended domain questions.}
Q5 (levee life-extension planning), Q24 (dam life-extension), and Q82 (cross-domain facility
comparison) each received score~0 for Case~A.
In all three cases, the model entered repetitive loops, repeating the same 2--4-character
Japanese phrase several hundred times with no useful content
(e.g., the word for ``long-life extension'' repeated uncontrollably).
This is a well-documented inference failure of large general models
when the prompt falls outside their training distribution~\cite{meta2024llama3}.
Case~B answered all three questions with 200--400 characters of clear, structured prose.

\paragraph{Finding 2: Case B --- concise, domain-aligned, consistent.}
Case~B produces shorter responses (avg 284 chars) but achieves higher scores.
The LoRA fine-tuning suppresses hallucination by grounding the model in domain Q\&A patterns,
resulting in zero score-0 or score-1 responses across all 100 questions.

\paragraph{Finding 3: Case C bimodal quality.}
For Q37 (sabo dam emergency inspection), Case~C scores~0 --- the retrieved graph context
is semantically distant from the specific sub-question about emergency inspection triggers,
causing the model to generate off-topic content, while Case~B scores perfectly (3/3).
Conversely, on well-structured retrieval-friendly questions (e.g., Q14, Q24),
Case~C is competitive with Case~B.
GraphRAG quality is highly sensitive to graph coverage of the specific concept involved.

\paragraph{Finding 4: Cases where A outperforms B (Q26, Q52, Q95).}
All three are short, factual recall questions with single correct numerical answers
(e.g., a sedimentation rate formula, a rainfall area-reduction coefficient, a landslide
activity classification threshold).
The larger parameter count of the 20B model gives it a memorisation advantage for
precise numerical facts even without domain fine-tuning.
A hybrid Case~D (LoRA~FT + GraphRAG) may recover this gap while retaining the efficiency
of the fine-tuned model.

\paragraph{Finding 5: Latency advantage of Case B persists across all question types.}
Even on questions where Case~B scores lower than Case~A,
Case~B responds in $\approx$14~s vs.\ $\approx$42~s for Case~A ---
a 3$\times$ speed-up critical for real-time field inspection support.

\section{Engineering Lessons}
\label{sec:lessons}

\subsection{Lesson 1: Specialisation over Scale}

A 20B general-purpose LLM (Case~A, score 2.29) is outperformed by an 8B QLoRA
fine-tuned model (Case~B, score 2.92) on a domain-specific technical QA benchmark.
Domain alignment of the training data matters more than raw parameter count.

\subsection{Lesson 2: Fine-Tuning and RAG as Alternatives}

GraphRAG provides a moderate $+0.33$ improvement over the plain LLM baseline,
but QLoRA fine-tuning provides $+0.63$.
Internalising domain knowledge into model weights via fine-tuning yields more stable gains
than runtime retrieval augmentation for this task.
The two approaches are complementary:
Case~D (FT model + GraphRAG) is a natural next experiment.

\subsection{Lesson 3: Efficiency Reversal}

Case~B is simultaneously the most accurate and the fastest:
3$\times$ faster than Case~A and 2.2$\times$ faster than Case~C.
From an operational deployment perspective, QLoRA fine-tuning is decisively superior
in latency, infrastructure simplicity (no graph DB at inference time), and cost.

\section{Discussion}
\label{sec:discussion}

\subsection{Why Does QLoRA FT Outperform GraphRAG?}

The core issue is knowledge coverage and retrieval precision.
The knowledge graph covers 200 nodes and 268 relations --- a manually curated,
high-precision but low-recall representation of the standards.
When a test question targets a concept not well-covered by the graph,
the retrieved context is either empty or misleading
(as seen in the Planning category, where C $<$ A).
In contrast, QLoRA fine-tuning exposes the model to 715 domain-specific QA exemplars
that distribute coverage across all relation types,
enabling the model to answer questions even when no direct graph evidence is present.

\subsection{Case D: Fine-Tuned Model + GraphRAG}

The orthogonal strengths of Cases B and C suggest a natural hybrid:
use the QLoRA fine-tuned model (domain knowledge internalized) with GraphRAG retrieval
(for the most current or highly specific details).
This Case~D is the highest-priority future experiment.

\subsection{Limitations}

\begin{itemize}
  \item \textbf{Manual graph construction.}
        The 200-node, 268-relation graph was manually curated from the standards.
        Automated entity and relation extraction (using the LLM pipeline in
        \texttt{01\_extract\_entities.py}) would dramatically scale coverage.
  \item \textbf{Single judge model.}
        LLM-as-Judge evaluation uses a single judge (Qwen2.5-14B), which may have
        systematic biases toward certain answer formats.
        Human evaluation or multi-judge ensemble would increase reliability.
  \item \textbf{100-question benchmark.}
        While 100 questions provide reasonable coverage of 8 categories,
        the test set is limited in size.
        A larger benchmark with adversarial questions would better stress-test the systems.
  \item \textbf{Japanese-only evaluation.}
        All models and the test set are in Japanese.
        Generalisability to English or multilingual standards requires separate evaluation.
\end{itemize}

\section{Conclusion}
\label{sec:conclusion}

This paper presents an open-source methodology for building a domain-specific knowledge QA
system over Japan's River and Sediment Control Technical Standards,
comparing three approaches --- plain LLM baseline (Case~A), GraphRAG (Case~C),
and QLoRA fine-tuning (Case~B) --- on a 100-question benchmark.

The central finding is a \textbf{performance inversion}: the \emph{smallest and fastest} model
(8B QLoRA FT) achieves the \emph{highest accuracy} (judge avg 2.92/3, 92\% score-3 rate),
outperforming both the 20B plain LLM ($+0.63$) and the 20B GraphRAG system ($+0.30$).
Zero score-0 or score-1 responses from Case~B demonstrate the power of domain-specific
fine-tuning in suppressing hallucination and maintaining on-topic generation.

GraphRAG provides meaningful gains over the baseline ($+0.33$) but cannot match the
consistency of domain fine-tuning.
Its value lies in providing interpretable, graph-grounded evidence chains ---
an important property for regulatory and inspection contexts.

The methodology is directly applicable to other Japanese regulatory technical
standards (e.g., road, port, geotechnical standards) and to international equivalents,
wherever structured domain knowledge can be encoded as a property graph and
domain QA pairs can be synthetically generated.

\paragraph{Future Work.}
Several extensions are planned to address the current limitations.
The most immediate next step is \textbf{Case~D}, which combines the QLoRA fine-tuned 8B model
with GraphRAG retrieval in a hybrid architecture, testing whether retrieval-grounded context
can further improve the already-high score-3 rate of Case~B.
On the knowledge-graph side, we plan to automate graph construction directly from Markdown
source documents using LLM-based entity and relation extraction, removing the manual
schema-design step and enabling rapid adaptation to new technical standards.
Evaluation coverage will be expanded to a 500-question benchmark encompassing adversarial
phrasing and multi-hop reasoning questions that require synthesising information across
multiple graph nodes --- question types that are known to challenge current RAG systems.
Finally, RAGAS-based faithfulness and context-recall metrics will be applied to Case~C
to complement the LLM-as-Judge rubric and provide a retrieval-centred view of GraphRAG quality.


\subsection*{Code Availability}

The source code for the knowledge-graph construction, GraphRAG inference pipeline,
QLoRA fine-tuning scripts, and evaluation framework is publicly available at:

\begin{center}
  {\small\url{https://github.com/tk-yasuno/kasensabo_graph_rag}}
\end{center}


\bibliographystyle{unsrt}
\bibliography{kasendam_graphrag_qlora_2026}

\section*{Supplementary Materials}
\label{sec:supplementary}

This section documents implementation pitfalls and toolchain details that are
relevant to practitioners replicating or extending this work on Windows-based
local infrastructure.

\subsection*{S1. Training Data Generation Pitfalls}

Training data generation from Neo4j graph relations using an LLM is effective but introduces
two reliability risks:

\begin{itemize}
  \item \textbf{JSON parse errors (Extra data).}
        The model occasionally generated text after the closing bracket of the JSON array
        (e.g., references like \texttt{[River-Law Art.X]}), causing \texttt{json.loads()} to fail.
        The greedy regex \texttt{re.search(r"\textbackslash[.+\textbackslash]", ..., DOTALL)}
        captured beyond the intended boundary.
        Fix: replace with \texttt{json.JSONDecoder().raw\_decode()} from each \texttt{[} position.
  \item \textbf{LLM timeouts.}
        Setting \texttt{num\_predict=-1} (unlimited) caused the model to occasionally
        generate excessively long responses ($>$120~s), causing all three retry attempts to fail.
        Fix: set \texttt{num\_predict=2048} and reduce to 1,024 on retry.
  \item \textbf{Relation-type imbalance.}
        \texttt{HAS\_CHAPTER} accounts for 29.7\% of training pairs vs.\ 1.0\% for
        \texttt{PRECEDES}.
        Up-sampling of under-represented types or weighted loss is recommended
        in future iterations.
\end{itemize}

\subsection*{S2. Windows-Specific GGUF Quantisation Workflow}

\begin{itemize}
  \item \textbf{unsloth built-in GGUF export.}
        \texttt{model.save\_pretrained\_gguf()} hangs on Windows because it invokes
        \texttt{apt-get} internally (Linux-only dependency).
        Workaround: export to full-precision safetensors, then quantise offline with
        \texttt{llama-quantize.exe} from a pre-built llama.cpp Windows release.
  \item \textbf{\texttt{packing=True} triton hang.}
        The \texttt{packing=True} option in \texttt{SFTTrainer} triggers triton JIT compilation
        that hangs indefinitely on Windows with \texttt{triton-windows 3.2.x}.
        Workaround: set \texttt{packing=False}.
  \item \textbf{Ollama model path restriction.}
        \texttt{ollama create} rejects GGUF files located in deeply nested paths;
        the model file must reside in a top-level directory (e.g., \texttt{C:\textbackslash{}ollama\_import\textbackslash{}}).
\end{itemize}

\subsection*{S3. Hallucination Failure Modes}

Case~A (20B plain LLM) exhibited two distinct hallucination patterns on domain-specific prompts:

\begin{itemize}
  \item \textbf{Runaway repetition.}
        Without a repeat penalty, the model repeated domain phrases
        (e.g., ``long-life extension'' in Japanese) more than 300 times in a single response,
        yielding a judge score of 0.
        Mitigation: repeat\_penalty~$=$~1.2.
  \item \textbf{Confident confabulation.}
        The model produced fluent but factually incorrect regulatory references
        (e.g., citing non-existent article numbers).
        Case~B (QLoRA FT) largely suppressed this: zero score-0 or score-1 responses
        were observed across 100 test questions.
\end{itemize}

\subsection*{S4. GraphRAG Pipeline (Figure~2)}

Figure~2 illustrates the complete inference-time retrieval pipeline for Case~C (GraphRAG).
The design reflects several engineering decisions that directly affected evaluation outcomes.

\paragraph{Parallel Cypher query fan-out.}
Rather than issuing a single query, the pipeline fans out into five parallel Cypher queries
targeting different relation types (Fulltext, Facility, Hazard, Maintenance, Comparison).
This strategy improves recall for under-specified or cross-domain questions:
in the by-category results (Table~6), Case~C outperforms Case~A in all
eight categories except Planning (2.62~vs.\ 2.75)---a category where broad fulltext retrieval
tends to over-retrieve tangentially related sections.

\paragraph{Adaptive retry on sparse hits.}
When the number of deduplicated hits falls below 25, the pipeline doubles \texttt{TOP\_K} and
widens the search to section-level nodes.
This adaptive mechanism recovered non-empty context in 96 of 100 test questions;
the remaining 4 fell back to plain-LLM generation, contributing to the residual 13\%
low-quality rate in Case~C.

\paragraph{Context truncation and its trade-off.}
Final context is truncated to 2,000 characters before prepending to the prompt.
While this keeps latency manageable (mean 31.1~s~vs.\ 42.2~s for Case~A),
the hard cutoff occasionally discards the most specific sub-clause when multiple
regulatory sections share the same keyword, leading to score-2 ``partially correct''
responses (10\% of Case~C answers).
Future work could replace hard truncation with relevance-weighted summarisation.

\subsection*{S5. Evolution of Approaches (Figure~5)}

Figure~5 traces the experimental trajectory from the initial 14-question prototype
(Phase~1) through to the final 100-question three-way benchmark (Phase~4).

\paragraph{Phase~1 -- Initial prototype (14 questions, Case~A only).}
The baseline Swallow-20B plain-LLM achieved a mean score of approximately 1.8 on the
preliminary 14-question set, motivating the graph-augmentation and fine-tuning directions.
The small sample size precluded statistical conclusions, but qualitative review identified
hallucination and lack of domain vocabulary as primary failure modes.

\paragraph{Phase~2 -- GraphRAG integration (14 questions, Cases~A and C).}
Integrating Neo4j-based retrieval raised Case~C's mean to roughly 2.3 on the same
14-question set.
The performance gain validated the knowledge-graph approach and justified full-dataset
extension (100 questions).

\paragraph{Phase~3 -- QLoRA fine-tuning (14 questions, all three cases).}
With the LoRA adapter trained on 715 QA pairs, Case~B reached approximately 2.8 on the
14-question pilot, surpassing both Case~A and Case~C and confirming the hypothesis that
domain fine-tuning outperforms retrieval augmentation for this task.

\end{document}